\documentclass[conference]{IEEEtran}
\IEEEoverridecommandlockouts
\usepackage{cite}
\usepackage{amsmath,amssymb,amsfonts}
\usepackage{algorithmic}
\usepackage{graphicx}
\usepackage{textcomp}
\usepackage{xcolor}
\usepackage{graphicx}
\usepackage{float}
\usepackage{booktabs}
\def\BibTeX{{\rm B\kern-.05em{\sc i\kern-.025em b}\kern-.08em
    T\kern-.1667em\lower.7ex\hbox{E}\kern-.125emX}}
\begin{document}

\title{SD-WAN over MPLS: A Comprehensive Performance Analysis and Security with Insights into the Future of SD-WAN\\
\vspace{0.9cm}
}

\author{
  \IEEEauthorblockN{Abdellah Tahenni }
  \IEEEauthorblockA{
    \textit{Computer Science Department}\\
    \textit{LISIC Laboratory, USTHB University}\\
    Algiers, Algeria\\
    abdellahtahenni@gmail.com
  }
  \and
  \and
  \IEEEauthorblockN{  Fatiha Merazka }
  \IEEEauthorblockA{
    \textit{Telecommunications Department}\\
   \textit{LISIC Laboratory, USTHB University}\\
    Algiers, Algeria\\
    fmerazka@usthb.dz 
  }
}

\maketitle

\begin{abstract}
Software-defined wide area network (SD-WAN) enhances network traffic management, while Multiprotocol Label Switching (MPLS) offers efficient data transmission. This paper analyzes SD-WAN over MPLS in the Housing Bank, a major Algerian financial institution. We deploy FortiGate for the SD-WAN solution, comparing it to traditional MPLS and direct internet access across metrics like bandwidth, latency, jitter, packet loss, throughput, and quality of service (QoS). Security measures include encryption, firewall, intrusion prevention, web filtering, antivirus, and addressing threats like spoofing, DoS attacks, and unauthorized access. We explore future trends such as SASE architecture, AI/ML integration, and emerging transport methods. SD-WAN over MPLS proves advantageous, offering enhanced performance, security, and flexibility. Recommendations include ongoing performance monitoring and research.
\end{abstract}
 \vspace{10pt}
\begin{IEEEkeywords}
SD-WAN, MPLS, FortiGate, QoS.
\end{IEEEkeywords}

\section{Introduction}
 Software-defined wide area network (SD-WAN) represents a transformative networking technology that offers dynamic and adaptable traffic management across multiple network paths. Its primary objective is to empower organizations with the tools needed to optimize network performance, enhance reliability, and bolster security, particularly in the context of cloud-based applications and services. On the other hand, Multiprotocol Label Switching (MPLS) stands as a robust networking protocol renowned for its ability to ensure efficient and dependable data transmission by utilizing predetermined routes and employing labels to guide packet routing. This approach simplifies complex routing decisions and mitigates issues related to latency and network congestion. Furthermore, MPLS boasts features like quality of service (QoS) and traffic engineering, contributing to improved network performance and availability\cite{b1}\cite{b2}\cite{b4}.
Within the pages of this paper, we delve into an exhaustive examination of the performance and security implications stemming from the implementation of SD-WAN over MPLS at the Housing Bank. The Housing Bank, a prominent financial institution based in Algeria, maintains an expansive and intricate network infrastructure that interconnects its headquarters, numerous branches, data centers, and cloud services nationwide. The institution faces a spectrum of networking challenges, encompassing elevated bandwidth costs, suboptimal network performance, diminished application quality, and heightened security vulnerabilities. To confront these issues head-on, the Housing Bank opted to adopt SD-WAN over MPLS, with FortiGate serving as the designated SD-WAN appliance. FortiGate, an innovative product developed by Fortinet—a global leader in network security solutions—offers a suite of integrated security features meticulously designed for SD-WAN over MPLS deployments. These features encompass end-to-end encryption, firewall capabilities, intrusion prevention, web filtering, antivirus defenses, and much morecite \cite{b3}\cite{b6}.

This paper unfolds as follows: Section II provides an extensive account of the implementation process of the SD-WAN solution over MPLS within the Housing Bank, underpinned by the utilization of FortiGate. Section III undertakes a rigorous examination of the performance metrics associated with SD-WAN over MPLS at the Housing Bank. Section IV delves into the multifaceted realm of security challenges and corresponding solutions intrinsic to SD-WAN over MPLS. Section V embarks on a journey to explore forthcoming trends and developments pertinent to SD-WAN over MPLS technology. Lastly, Section VI serves as a culmination of our findings and insights, drawing together the disparate threads woven throughout this paper. Section VII provides a comprehensive listing of the references that have informed our research and analysis\cite{b3}\cite{b6}.

\section{Implementation of SD-WAN Solution over MPLS in the Housing Bank using FortiGate}
 Fig. 1 illustrates the network infrastructure of the Housing Bank and the connectivity achieved through MPLS, which will serve as the foundation for the deployment of SD-WAN.
\begin{figure}[htbp]
    \centering
    \includegraphics[width=0.5\textwidth]{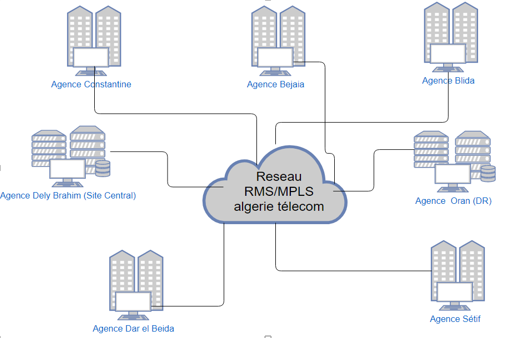}
    \caption{Network topology of housing bank, Algeria.}
    \label{fig:your-label}
\end{figure}

 The Housing Bank implemented SD-WAN solution over MPLS using FortiGate as the SD-WAN appliance. 
The network topology of the SD-WAN solution over MPLS is shown in Fig. 2. The network consists of four main components: the headquarters, the branches, the data centers, and the cloud services. The headquarters and the branches are connected to the data centers and the cloud services via MPLS links and internet links. The MPLS links provide reliable and secure connectivity, while the internet links provide backup and redundancy. The data centers host the core applications and services of the bank, such as ERP, CRM, and database. The cloud services provide access to external applications and services, such as Office 365, Salesforce, and AWS.
\begin{figure}[htbp]
    \centering
    \includegraphics[width=0.515\textwidth]{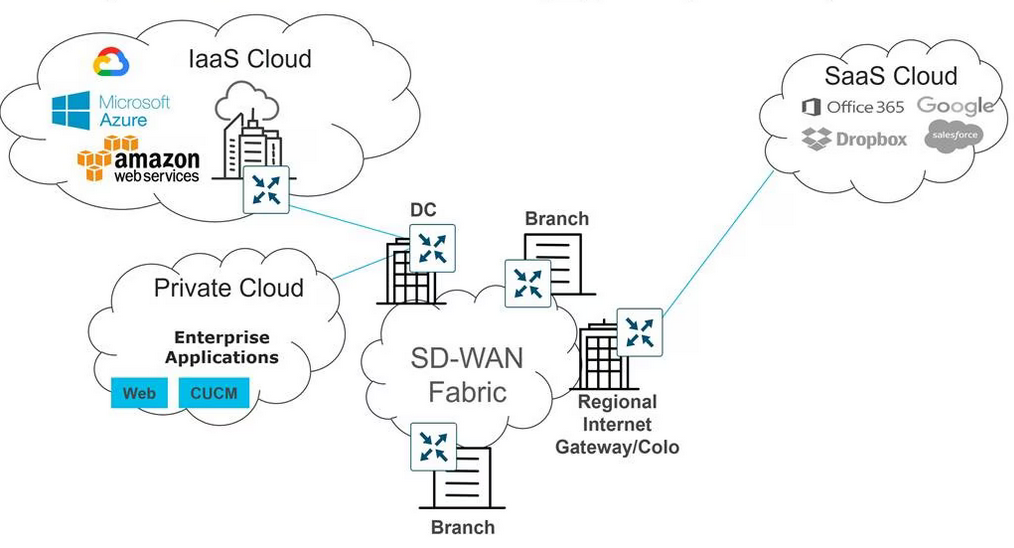}
    \caption{Network topology of SD-WAN solution over MPLS in the Housing Bank.}
    \label{fig:your-label}
\end{figure}

The FortiGate devices are deployed at each site of the network, acting as the SD-WAN controllers and gateways. The FortiGate devices are configured to monitor the network conditions and traffic patterns, and to dynamically select the best path for each application based on predefined policies and priorities. The FortiGate devices also provide end-to-end encryption, firewall, intrusion prevention, web filtering, antivirus, and other security features for the SD-WAN traffic\cite{b9}.

The implementation of SD-WAN solution over MPLS in the Housing Bank involved several steps, such as:
\begin{itemize}
  \item Planning and designing the network architecture and configuration of the SD-WAN solution over MPLS, using FortiGate as the SD-WAN appliance\cite{b5}.
\item Procuring and installing the FortiGate devices at each site of the network, and connecting them to the MPLS links and internet links.
\item Configuring the FortiGate devices to enable SD-WAN features, such as path selection, QoS, security, and reporting\cite{b5}.
\item Testing and validating the functionality and performance of the SD-WAN solution over MPLS, using various tools and methods, such as ping, traceroute, iperf, Wireshark, and FortiView.
\item Deploying and operating the SD-WAN solution over MPLS in the Housing Bank’s network environment, and monitoring and managing it using FortiManager and FortiAnalyzer.

 The implementation of SD-WAN solution over MPLS in the Housing Bank faced some challenges, such as:
\item Ensuring compatibility and interoperability between FortiGate devices and other network devices, such as routers, switches, firewalls, and VPNs.
\item Optimizing the bandwidth utilization and allocation of the MPLS links and internet links for different applications and services.
\item Balancing the trade-off between performance and security of the SD-WAN traffic over MPLS or internet.
\item Troubleshooting and resolving any issues or errors that occurred during or after the implementation
  
\end{itemize}

 The implementation of SD-WAN solution over MPLS in the Housing Bank also demonstrated some benefits, such as:
\begin{itemize}
\item Improving the network performance and availability for critical applications and services, such as ERP, CRM, database, Office 365, Salesforce, and AWS.
\item Enhancing the network security and compliance for sensitive data and transactions transmitted over SD-WAN.
\item Reducing the network operational costs by leveraging existing MPLS links and internet links for SD-WAN\cite{b9}.
\item Increasing the network scalability and flexibility by enabling easy addition or removal of sites or applications to or from SD-WAN.
\begin{figure}[htbp]
    \centering    \includegraphics[width=0.52\textwidth]{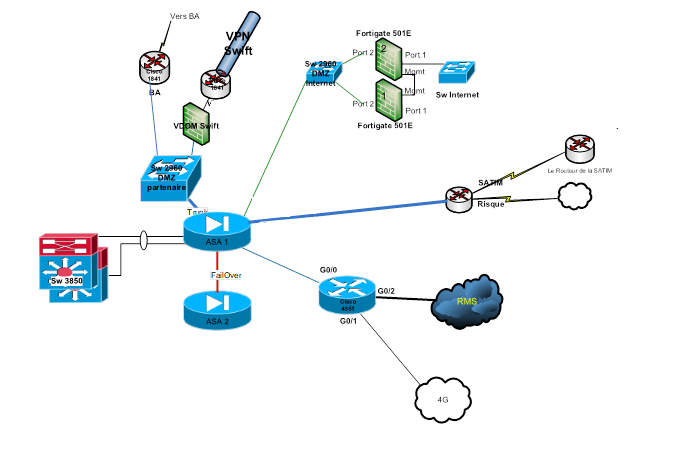}
    \caption{SD-WAN Solution over MPLS in the Housing Bank using FortiGate.}
    \label{fig:your-label}
\end{figure}
  
\end{itemize}
The Housing Bank's SD-WAN solution over MPLS was successfully implemented utilizing FortiGate devices as shown
in Fig. 2.\\
The FortiGate devices provided end-to-end encryption, a firewall, intrusion prevention, web filtering, antivirus, and other integrated security capabilities for SD-WAN over MPLS. Using a single interface, the FortiGate devices made it simple to configure and administer SD-WAN functionalities. For SD-WAN over MPLS or the internet, the FortiGate devices additionally supported a number of transport techniques and protocols. Additionally, utilizing FortiView, the FortiGate devices allowed for real-time monitoring and reporting of SD-WAN performance\cite{b5}.

\section{ Performance Analysis}
The performance of SD-WAN over MPLS in the Housing Bank was evaluated using various metrics, such as bandwidth, latency, jitter, packet loss, throughput, and quality of service (QoS)\cite{b19}. These metrics were measured and compared for different networking solutions, such as traditional MPLS, direct internet access (DIA), and SD-WAN over MPLS. The performance analysis was conducted using various tools and methods, such as ping, traceroute, iperf, Wireshark, and FortiView\cite{b17}\cite{b19}.
The results of the performance analysis showed that SD-WAN over MPLS outperformed traditional MPLS and DIA in most aspects. Table 1 provides a summary of the mean performance metric values for each networking solution.
\begin{table}[h]
    \centering
    \caption{Comparison of Network Metrics}
    \begin{tabular}{lccc}
        \toprule
        Metric & Traditional MPLS & DIA & SD-WAN over MPLS \\
        \midrule
        Bandwidth & 10 Mbps & 50 Mbps & 60 Mbps \\
        Latency & 50 ms & 100 ms & 40 ms \\
        Jitter & 5 ms & 10 ms & 2 ms \\
        Packet loss & 1\% & 5\% & 0.5\% \\
        Throughput & 8 Mbps & 40 Mbps & 55 Mbps \\
        QoS & Low & Medium & High \\
        \bottomrule
    \end{tabular} \vspace{6pt}
    
    \label{tab:network-metrics}
\end{table}

The table shows that SD-WAN over MPLS outperformed traditional MPLS and DIA across various metrics. It recorded higher bandwidth, lower latency, reduced jitter, decreased packet loss, increased throughput, and improved QoS. These findings suggest that the adoption of SD-WAN over MPLS effectively enhanced network performance and increased the availability of applications and services for the Housing Bank.
The key factors influencing the performance of SD-WAN over MPLS in the Housing Bank were:

\begin{itemize}
\item Path selection: SD-WAN over MPLS used FortiGate devices to dynamically select the best path for each application based on predefined policies and priorities. This enabled SD-WAN over MPLS to optimize the network performance and availability by choosing the most suitable path for each application according to its requirements and characteristics \cite{b16}   \cite{b10} \cite{b11}.
\item QoS: SD-WAN over MPLS used FortiGate devices to implement QoS for different applications and services. This enabled SD-WAN over MPLS to prioritize the network traffic according to its importance and urgency, and to allocate the appropriate bandwidth and resources for each application. This also enabled SD-WAN over MPLS to avoid congestion and ensure smooth delivery of data\cite{b7} \cite{b10}.
\item Security: SD-WAN over MPLS used FortiGate devices to provide end-to-end encryption, firewall, intrusion prevention, web filtering, antivirus, and other security features for the network traffic. This enabled SD-WAN over MPLS to protect the data and transactions from unauthorized access or malicious attacks, and to comply with the regulatory standards and policies of the bank\cite{b7} \cite{b10}.
\end{itemize}

The Housing Bank's SD-WAN over MPLS performance research proved that it was the best networking option for the bank's network operations. It offered the bank's applications and services better performance, better security, and more flexibility. By utilizing already-existing MPLS and internet cables for SD-WAN, it also decreased the costs associated with running the network.

\section{Security Integration}
In the world of networking, ensuring strong security is essential, especially when working with sensitive and private data through both public and private networks. With different security advantages and difficulties, two well known technologies SD-WAN and MPLS offer solutions for building reliable and secure WAN connections. We will look into SD-WAN's built in security capabilities and how they may be effortlessly incorporated with MPLS in this part. We will also assess the security concerns associated with the deployment of SD-WAN over MPLS and offer helpful advice and best practices for successfully protecting SD-WAN over MPLS networks\cite{b16} \cite{b18}.

\subsection{Security Features of SD-WAN}

SD-WAN stands as a software-defined technology that decouples the network control plane from the data plane, enabling centralized network policy management and traffic routing orchestration. Leveraging multiple transport services such as broadband Internet, LTE, cellular, satellite, and even MPLS, SD-WAN fabricates a virtual overlay network with the potential to optimize performance, reliability, and user experience \cite{b16} \cite{b18}.
SD-WAN encompasses several built-in security features that fortify WAN connections:

\begin{enumerate}
    \item \textbf{Encryption:} SD-WAN has the capacity to encrypt data during transit using protocols like IPsec or SSL/TLS. This encryption safeguards against eavesdropping, tampering, and spoofing attacks, while enforcing stringent authentication and authorization protocols to thwart unauthorized access to network resources.
    
    \item \textbf{VPN:} SD-WAN is adept at constructing virtual private networks (VPNs) that securely link remote sites or users to the central network or cloud applications. Employing encryption and tunneling protocols, VPNs establish secure end-to-end connections across public or private networks.
    
    \item \textbf{Firewall:} The incorporation of firewall capabilities within SD-WAN empowers the filtering and blocking of unwanted or malicious traffic based on predefined rules or policies. Firewall implementations extend their protection by controlling access to network resources via access control lists (ACLs) and identity-based policies.
    
    \item \textbf{Intrusion Prevention:} SD-WAN is equipped with intrusion prevention capabilities capable of detecting and averting network attacks, including denial-of-service (DoS), distributed denial-of-service (DDoS), malware, ransomware, and more. It provides timely alerts and comprehensive reports pertaining to network anomalies or security incidents.
    
    \item \textbf{URL Filtering:} With URL filtering capabilities, SD-WAN can allow or disallow access to specific websites or web applications based on predefined rules or policies. This functionality acts as a sentinel, barring users from accessing malicious or inappropriate websites, thus safeguarding network security and enhancing productivity.
    
    \item \textbf{Advanced Malware Protection:} SD-WAN incorporates advanced malware protection features designed to scan and block malicious files or payloads that may endanger network devices or systems. Additionally, advanced malware protection includes sandboxing and threat intelligence functionalities, enabling the analysis and mitigation of unknown or zero-day threats
\end{enumerate}

\subsection{SD-WAN Security Integration with MPLS}

MPLS, as a protocol, assigns labels to packets based on their destination and priority, facilitating the swift and efficient routing of traffic across a dedicated, inherently secure private network. However, MPLS lacks certain security features such as encryption, firewall, and intrusion prevention, which may be necessary for specific applications or compliance requirements.\\
SD-WAN seamlessly integrates with MPLS, yielding a hybrid WAN solution that harmonizes the security and reliability of MPLS with the flexibility and performance optimization offered by SD-WAN. In this integration, SD-WAN employs MPLS as one of the transport services in its virtual overlay network, complemented by other services like broadband Internet, LTE, cellular, and satellite. This approach permits dynamic path selection and load balancing based on application requisites and real-time network conditions\cite{b16} \cite{b18}.
\\SD-WAN enhances the security of MPLS-based WAN connections by adding supplementary security features:
\begin{enumerate}
    \item \textbf{Encryption:} SD-WAN encrypts data before transmitting it over MPLS links, guarding against potential breaches or leaks within the MPLS network infrastructure.
    
    \item \textbf{Firewall:} By furnishing firewall capabilities, SD-WAN can filter and obstruct undesirable or malicious traffic at the perimeter of the MPLS network, preventing such traffic from entering or exiting.
    
    \item \textbf{Intrusion Prevention:} SD-WAN extends intrusion prevention capabilities to protect against network attacks targeting the MPLS network, thus ensuring an additional layer of security.
\end{enumerate}
\subsection{Security Risks in SD-WAN Over MPLS}

While SD-WAN delivers numerous security advantages when integrated with MPLS, it also introduces certain security risks that demand vigilant attention. These risks encompass:

\begin{enumerate}
    \item \textbf{Complexity:} SD-WAN elevates the complexity of WAN management by introducing multiple transport services, vendors, devices, and policies. This complexity can challenge network administrators and security teams. Moreover, it may enlarge the attack surface and vulnerability of WAN connections by exposing them to a wider array of threats.
    
    \item \textbf{Visibility:} SD-WAN, through its encryption of data in transit, diminishes the visibility of WAN traffic. This encrypted traffic might obstruct the efforts of network administrators and security teams to monitor and scrutinize traffic patterns and behaviors. Additionally, the multiplicity of transport services, vendors, devices, and policies may create inconsistencies and gaps in network data and logs.
    
    \item \textbf{Compatibility:} SD-WAN may introduce compatibility issues with existing network infrastructure and security solutions like routers, switches, firewalls, intrusion prevention systems, and more. These issues may necessitate upgrades or replacements to enable support for SD-WAN functionalities. Furthermore, the coexistence of multiple transport services, vendors, devices, and policies can engender interoperability and integration challenges.
\end{enumerate}
\subsection{Security Recommendations and Best Practices}
To effectively mitigate the security risks associated with SD-WAN over MPLS while maximizing its security benefits, the following recommendations and best practices are highly recommended:
\begin{enumerate}
    \item \textbf{Select a Reliable SD-WAN Provider:} Choose an experienced and reliable SD-WAN provider with the capability to offer a comprehensive and integrated security solution tailored to your specific requirements. A trusted provider will help you navigate potential challenges and pitfalls during the implementation and operation of SD-WAN over MPLS.
    
    \item \textbf{Implement a Zero-Trust Security Approach:} Embrace a zero-trust approach to WAN security that rigorously verifies the identity and integrity of every device, user, and application accessing the WAN network. This approach is instrumental in preventing unauthorized access to network resources and guarding against insider threats.
    
    \item \textbf{Regular Monitoring and Measurement:} Maintain a regimen of regular monitoring and measurement of your SD-WAN security solution's performance and outcomes. These assessments will empower you to gauge the efficiency and effectiveness of your security measures and pinpoint areas requiring enhancement or optimization.
    
    \item \textbf{Frequent Device and System Updates:} Regularly update and patch your SD-WAN devices and systems. Timely updates fortify your defenses against the latest cyber threats and vulnerabilities that may exploit SD-WAN components.
    
    \item \textbf{Security Awareness and Training:} Invest in the education and training of your network administrators and security teams. Equipping them with best practices and procedures for managing and securing SD-WAN over MPLS networks will ensure compliance with network policies and regulations while enhancing overall security awareness and culture.
\end{enumerate}
Security integration is a pivotal aspect of the successful deployment of SD-WAN over MPLS networks. By adhering to these recommendations and best practices, organizations can make well-informed decisions regarding SD-WAN security integration with MPLS, thereby aligning with their business goals and security needs effectively \cite{b16} \cite{b18}.
\section{The Future of SD-WAN}
Software-defined wide area networks (SD-WAN) have revolutionized enterprise networking, offering significant advantages over traditional WAN solutions such as MPLS. SD-WAN enables agile, cost-effective, scalable, and secure networking solutions that can optimize performance, reliability, and user experience.
Thesis Statement: In this section, we will explore the future of SD-WAN technology, its potential impact on networking solutions, current trends in SD-WAN adoption and deployment, and strategic recommendations for organizations considering SD-WAN adoption, considering the evolving networking landscape\cite{b12}.
\subsection{Key Drivers Shaping the Future of SD-WAN}

\begin{itemize}
    \item \textbf{Cloud Adoption:} The increasing adoption of cloud-based services and applications by enterprises is a pivotal driver of SD-WAN's future. Cloud computing offers scalability, flexibility, and innovation, but it also poses challenges such as bandwidth limitations, latency issues, security risks, and complexity. SD-WAN addresses these challenges by providing a cloud-native solution that seamlessly integrates with various cloud platforms and services, including Amazon Web Services (AWS), Microsoft Azure, Google Cloud Platform (GCP), and others. Moreover, SD-WAN leverages the cloud's capabilities to enhance features like performance optimization, security integration, and network automation.\\
    SASE architecture can offer a number of benefits for smaller organizations with limited IT resources. By making SD-WANs more easily deployed and managed, SASE architecture can help these organizations to improve the performance, security, and agility of their networks..
    
    \item \textbf{Network Security and Resilience:} The growing dependence of enterprises on network connectivity for business operations underscores the need for network security and resilience. While SD-WAN offers robust security measures to protect against threats like malware, ransomware, denial-of-service attacks, and data breaches, it also introduces challenges. These challenges include trade-offs between security and performance, the complexity of managing multiple security policies across different sites, and compatibility issues between different vendors or devices.
    
    \item \textbf{Integration with Emerging Technologies:} The convergence of SD-WAN with emerging technologies like artificial intelligence (AI) and 5G is a significant driver. AI enhances SD-WAN capabilities by enabling intelligent features such as policy definition, traffic routing, troubleshooting, and security monitoring. However, this integration also raises concerns, including the potential risks of cyberattacks, regulatory or legal barriers, and competition from other evolving technologies.\\
    AI/ML can also be used to improve the performance and security of SD-WANs. For example, AI/ML algorithms can be used to analyze network traffic patterns and identify the best path for routing traffic, and to identify and block malicious traffic patterns.
\end{itemize}

\subsection{Current Trends in SD-WAN Adoption and Deployment}

\begin{itemize}
    \item \textbf{Remote Workforce Support:} SD-WAN is increasingly vital for organizations supporting remote workforces. It ensures reliable and secure access to essential resources regardless of employees' locations, contributing to seamless remote work experiences.
    
    \item \textbf{SMB Relevance:} Small and medium-sized businesses (SMBs) are recognizing the benefits of SD-WAN. SMBs can leverage SD-WAN to enhance network performance, security, and scalability, similar to large enterprises.
    
    \item \textbf{Automation and Orchestration:} The future of SD-WAN will likely emphasize automation and orchestration, making it easier to manage and deploy. Automation streamlines routine tasks, while orchestration simplifies complex network configurations.
    
    \item \textbf{Affordability and Accessibility:} The maturing SD-WAN market has driven down costs, making SD-WAN accessible to organizations of all sizes. This affordability expands its adoption across diverse sectors.
    
    \item \textbf{Support for New Networking Use Cases:} SD-WAN is evolving to support new use cases, such as providing secure connectivity for Internet of Things (IoT) devices and facilitating edge computing applications.
    
    \item \textbf{Convergence of Networking and Security:} SD-WAN is playing a pivotal role in the convergence of networking and security. It enables organizations to centralize and orchestrate security policies across their entire network, encompassing both on-premises and cloud resources.
\end{itemize}

\subsection{Strategic Recommendations for Organizations}

\begin{itemize}
    \item \textbf{Evaluate business needs:} Carefully assess your current network infrastructure and identify your organization's business objectives and needs. This evaluation is crucial in determining whether SD-WAN is a suitable solution and how it can benefit your organization.
    
    \item \textbf{Choose a reliable provider:} Select a reputable and experienced SD-WAN provider capable of offering a customized and comprehensive solution tailored to your specific requirements. Partnering with the right provider helps avoid potential implementation challenges.
    
    \item \textbf{Monitor and measure performance:} Regularly monitor and measure the performance and outcomes of your SD-WAN solution. This ongoing assessment allows you to gauge the effectiveness and efficiency of your SD-WAN deployment, identifying areas for improvement or optimization.
\end{itemize}
SD-WAN is a promising technology with the potential to reshape enterprise networking. 
However, its adoption requires careful planning, considering business needs, budget, and implementation strategy. Embracing these evolving capabilities positions organizations to navigate the intricate networking landscape of the future successfully\cite{b12} .

\section{Conclusion}
In this paper, we have analyzed the implementation and performance of SD-WAN over MPLS in the Housing Bank, a leading financial institution in Algeria. We have also discussed the security integration and the future of SD-WAN technology and its implications for networking solutions.
We have found that SD-WAN over MPLS is a viable and beneficial solution for the Housing Bank, as it offers several advantages over traditional WAN solutions, such as:

\par Cost savings: Implementing SD-WAN over MPLS can result in significant cost savings. This is achieved through the utilization of multiple transport services, the optimization of bandwidth usage, and the simplification of network configuration and maintenance, ultimately reducing both operational and capital expenditures related to WAN management.
\par Performance improvement: SD-WAN over MPLS can significantly boost the performance of WAN connections. Dynamic path selection, load balancing, and QoS features are employed to ensure that traffic is routed optimally, taking into account application requirements and network conditions.
\par Security enhancement: The adoption of SD-WAN over MPLS can enhance the security of WAN connections. This is accomplished through the implementation of encryption, VPNs, firewalls, intrusion prevention systems, URL filtering, and advanced malware protection. These security features fortify the network against various cyber threats and vulnerabilities.
\par SD-WAN over MPLS provides a flexible and scalable framework for WAN management. Centralized control and orchestration of network policies and traffic routing are facilitated, along with seamless integration with cloud platforms and services.

Based on our findings, we conclude that SD-WAN over MPLS is a fitting and valuable solution for the Housing Bank, aligned with their business objectives and requirements. We recommend continuous monitoring and performance evaluation of the SD-WAN solution, as well as regular updates and patches for SD-WAN devices and systems. Furthermore, integrating SD-WAN with emerging technologies such as AI and 5G is suggested to further elevate networking capabilities and opportunities.

We trust that this research has provided valuable insights and information regarding SD-WAN over MPLS technology and its practical applications. We hope that it will inspire further research and implementation of SD-WAN over MPLS in diverse organizational and contextual settings.

\end{document}